\documentclass[runningheads]{llncs}
\usepackage{graphicx}
\usepackage{amsmath}

\newcommand{\LL}{\ensuremath{\mathbf{L}}}

\begin{document}

\renewcommand{\arraystretch}{1.4}
\setlength{\tabcolsep}{0.3em}

\title{Robust Semantic Segmentation of Brain Tumor Regions from 3D MRIs}
\titlerunning{Robust Semantic Segmentation of Brain Tumor Regions from 3D MRIs}
\author{Andriy Myronenko, Ali Hatamizadeh}
\authorrunning{A. Myronenko}
\institute{NVIDIA, Santa Clara, CA \\ \email{amyronenko@nvidia.com, ahatamizadeh@nvidia.com} }
\maketitle              
\begin{abstract}
Multimodal brain tumor segmentation challenge (BraTS) brings together researchers to improve automated methods for 3D MRI brain tumor segmentation.  Tumor segmentation is one of the fundamental vision tasks necessary for diagnosis and treatment planning of the disease. Previous years winning methods were all deep-learning based, thanks to the advent of modern GPUs, which allow fast optimization of deep convolutional neural network architectures. 
In this work, we explore best practices of 3D semantic segmentation, including conventional encoder-decoder architecture, as well combined loss functions, in attempt to further improve the segmentation accuracy. We evaluate the method on BraTS 2019 challenge. 
\end{abstract}

\section{Introduction}

Brain tumors are categorized into primary and secondary tumor types.  Primary brain tumors 	 originate from brain cells, whereas  secondary tumors metastasize into the brain from other organs. The most common type of primary brain tumors are gliomas, which arise from brain glial cells.  Gliomas can be of low-grade  (LGG) and high-grade (HGG) subtypes. High grade gliomas are an aggressive  type of malignant brain tumor that grow rapidly, usually require surgery and radiotherapy and have poor survival prognosis. Magnetic Resonance Imaging (MRI) is a key diagnostic tool for brain tumor analysis, monitoring and surgery planning. Usually, several complimentary 3D MRI modalities are acquired - such as T1, T1 with contrast agent (T1c), T2 and Fluid Attenuation Inversion Recover (FLAIR) - to emphasize different tissue properties and areas of tumor spread.  For example the contrast agent, usually gadolinium, emphasizes hyperactive tumor subregions in T1c MRI modality.  

Automated segmentation of 3D brain tumors can save physicians time and provide an accurate reproducible solution for further tumor analysis and monitoring. Recently, deep learning based segmentation techniques surpassed traditional computer vision methods for dense semantic segmentation. Convolutional neural networks (CNN)  are able to learn from examples and demonstrate state-of-the-art segmentation accuracy both in 2D natural images~\cite{deeplabv3plus2018,hatamizadeh2019end} and in 3D medical image modalities~\cite{Milletari16}. 

\begin{figure}[t] 
 	\centering
 	\includegraphics[clip=true, trim=0pt 0pt 0pt 0pt, width=0.9\textwidth]{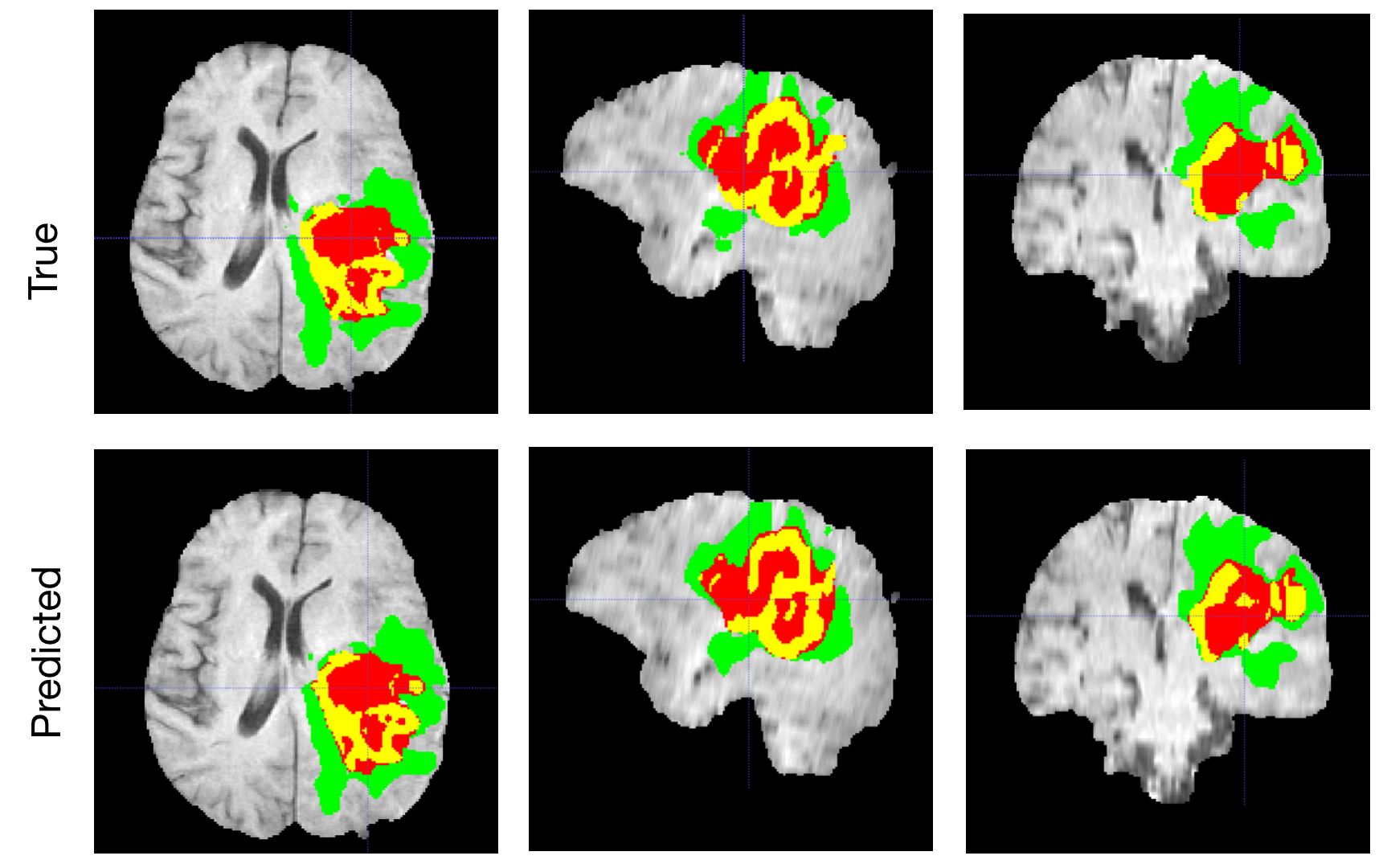}
 	\caption{A typical segmentation example with true and predicted labels overlaid over T1c MRI axial, sagittal and coronal slices.  The whole tumor (WT) class includes all visible labels (a union of green, yellow and red labels), the tumor core (TC) class is a union of red and yellow, and the enhancing tumor core (ET) class is shown in yellow (a hyperactive tumor part). The predicted segmentation results match the ground truth well.}
 	\label{fig:seg}
 \end{figure}

Multimodal Brain Tumor Segmentation Challenge (BraTS) aims to evaluate state-of-the-art methods for the segmentation of brain tumors by providing a 3D MRI dataset with ground truth tumor segmentation labels annotated by physicians~\cite{BratsAll2018,brats1,brats2,brats3,brats4}. This year, BraTS 2019 training dataset included 335 cases, each with four 3D MRI modalities (T1, T1c, T2 and FLAIR) rigidly aligned, resampled to 1x1x1 mm isotropic resolution and skull-stripped. The input image size is 240x240x155. The data were collected from multiple institutions, using various MRI scanners. Annotations include 3 tumor subregions: the enhancing tumor, the peritumoral edema, and the necrotic and non-enhancing tumor core.  The annotations were combined into 3 nested subregions: whole tumor (WT), tumor core (TC) and enhancing tumor (ET), as shown in Figure~\ref{fig:seg}. Two additional datasets without the ground truth labels were provided for validation and testing. These datasets required participants to upload the segmentation masks to the organizers' server for evaluations. The validation dataset (125 cases) allowed multiple submissions and was designed for intermediate evaluations. The testing dataset allowed only a single submission, and is used to calculate the final challenge ranking.  

In this work, we describe our semantic segmentation approach for volumetric 3D brain tumor segmentation from multimodal 3D MRIs and participate in BraTS 2019 challenge.


\section{Related work}
\label{sec:relatedwork}

Previous year, BraTS 2018 top submissions included  Myronenko~\cite{Myronenko18},  Isensee et al.~\cite{Isensee18brats}, McKinly et al.~\cite{McKinley18brats} and Zhou et al.~\cite{Zhou18brats}. In our previous work~\cite{Myronenko18}, we explored how an additional decoder for a secondary task get impose additional structure on the network. Isensee et al.~\cite{Isensee18brats} demonstrated that a generic U-net architecture with a few  minor modifications is enough to achieve competitive performance. McKinly et al.~\cite{McKinley18brats} proposed a segmentation CNN in which a DenseNet~\cite{huang2017densely} structure with dilated convolutions was embedded in U-net-like network. Finally, Zhou et al.~\cite{Zhou18brats} proposed to use an ensemble of different networks: taking into account multi-scale context information,  segmenting 3 tumor subregions in cascade with a shared backbone weights and adding an attention block.  

Here, we generally follow the previous year submission~\cite{Myronenko18}, but instead of secondary task decoder we explore various architecture design choices and complimentary loss functions. We also utilize multi-gpu systems for data parallelism to be able to use larger batch sizes.  


\section{Methods}

\label{sec:methods}

Our segmentation approach generally follows~\cite{Myronenko18} with encoder-decoder based CNN architecture.

\subsection{Encoder part}
The encoder part uses  ResNet~\cite{He16}  blocks, where each block consists of two convolutions with normalization and ReLU, followed by additive identity skip connection. For normalization, we experimented with Group Normalization (GN)~\cite{Wu18}, Instance Normalization~\cite{Instance16} and Batch Normalization~\cite{Ioffe15}.  We follow a common CNN approach to progressively downsize image dimensions by 2 and simultaneously increase feature size by 2.  For downsizing we use strided convolutions.   All convolutions are 3x3x3 with initial number of filters equal to 32.  
The encoder part structure is shown in Table~\ref{tab:encoder}. 
The encoder endpoint has size 256x20x24x16, and is 8 times spatially smaller than the input image. We decided against further downsizing to preserve more spatial content. 

\begin{table}
	\centering
	\caption{Encoder structure, where GN stands for group normalization (with group size of 8), Conv - 3x3x3 convolution, AddId - addition of identity/skip connection. Repeat column shows the number of repetitions of the block. We refer to the final output of the encoder, as the encoder endpoint}
	\label{tab:encoder}
	\begin{tabular}{|l|c|c|c|}
		 \hline
		Name & Ops & Repeat &Output size    \\ \hline
		Input & &  &4x160x192x128    \\
		InitConv & Conv & 1 & 32x160x192x128    \\
		EncoderBlock0 & GN,ReLU,Conv,GN,ReLU,Conv, AddId & 1 &32x160x192x128    \\
		EncoderDown1 & Conv stride 2 & 1&64x80x96x64    \\
		EncoderBlock1 & GN,ReLU,Conv,GN,ReLU,Conv, AddId& 2 &64x80x96x64    \\
		EncoderDown2 & Conv stride 2& 1&128x40x48x32    \\
		EncoderBlock2 & GN,ReLU,Conv,GN,ReLU,Conv, AddId& 2 &128x40x48x32    \\
		EncoderDown3 & Conv stride 2& 1 &256x20x24x16    \\
		EncoderBlock3 & GN,ReLU,Conv,GN,ReLU,Conv, AddId& 4&256x20x24x16   \\
		\hline
	\end{tabular}
\end{table}

 \subsection{Decoder part}

The decoder structure is similar to the encoder one, but with a single block per each spatial level. Each decoder level begins with upsizing: reducing the number of features  by a factor of 2 (using 1x1x1 convolutions) and doubling the spatial dimension (using 3D bilinear upsampling),  followed by an addition of encoder output of the equivalent spatial level. The end of the decoder has the same spatial size as the original image, and the number of features equal to the initial input feature size, followed by 1x1x1 convolution into 3 channels and a sigmoid function.  The decoder structure is shown in Table~\ref{tab:decoder}.

\begin{table}
	\centering
	\caption{Decoder structure, where GN stands for group normalization (with group size of 8), Conv - 3x3x3 convolution, Conv1 - 1x1x1 convolution, AddId - addition of identity/skip connection, UpLinear - 3D linear spatial upsampling }
	\label{tab:decoder}
	\begin{tabular}{|l|c|c|c|}
		\hline
		Name & Ops & Repeat &Output size    \\ \hline
		DecoderUp2 & Conv1, UpLinear, +EncoderBlock2 &  1 &128x40x48x32    \\
		DecoderBlock2 & GN,ReLU,Conv,GN,ReLU,Conv, AddId & 1 & 128x40x48x32    \\
		DecoderUp1 & Conv1, UpLinear, +EncoderBlock1 &  1 &64x80x96x64    \\
		DecoderBlock1 & GN,ReLU,Conv,GN,ReLU,Conv, AddId & 1 & 64x80x96x64   \\
		DecoderUp0 & Conv1, UpLinear, +EncoderBlock0 &  1 &32x160x192x128    \\
		DecoderBlock0 & GN,ReLU,Conv,GN,ReLU,Conv, AddId & 1 & 32x160x192x128   \\
		DecoderEnd & Conv1, Sigmoid &  1 &1x160x192x144    \\
		\hline
	\end{tabular}
\end{table}

 \subsection{Loss}
We use a hybrid loss function that consists of the following terms:
  \begin{equation}
 \label{eq:loss}
 \LL =\LL_{dice} + \LL_{focal} + \LL_{acl}   
 \end{equation}
$\LL_{dice}$ is a soft dice loss~\cite{Milletari16} applied to the  decoder output $p_{pred}$ to match the segmentation mask $p_{true}$:
   \begin{equation}
  \label{eq:dice}
  \LL_{dice}  = 1 - \frac{2*\sum p_{true} * p_{pred} }{\sum p_{true}^2 + \sum p_{pred}^2 + \epsilon}   
  \end{equation}
where summation is voxel-wise, and $\epsilon$ is a small constant to avoid zero division. Since the output of the segmentation decoder has 3 channels (predictions for each tumor subregion), we simply add the three dice loss functions together. 
$\LL_{acl}$ is the 3D extension of supervised active contour loss \cite{chen2019learning} that consists of volumetric and length terms:

   \begin{equation}
  \label{eq:acl}
  \LL_{acl}  = \LL_{vol} + \LL_{length}
  \end{equation}

in which :

   \begin{equation}
  \label{eq:acl_volume}
  \LL_{vol}  = \mid\sum p_{pred}(c_{1}-p_{true})^2\mid+\mid\sum (1-p_{pred})(c_{2}-p_{true})^2\mid 
  \end{equation}

   \begin{equation}
  \label{eq:acl_length}
  \LL_{length}  =\sum \sqrt{\mid(\nabla p_{pred,x})^2+(\nabla p_{pred,y})^2+(\nabla p_{pred,z})^2\mid + \epsilon}
  \end{equation}

Where $c_{1}$ and $c_{2}$ represent the energy of the foreground and 
background.
$\LL_{focal}$ is a focal loss function \cite{lin2017focal} defined as: 


\begin{equation}
  \label{eq:focal}
  \LL_{focal}  =-\frac{1}{N}\sum(1 - p_{pred})^{\gamma} p_{true} \log\left(p_{pred}+\epsilon\right)
  \end{equation}

Where $N$ is the total number of voxels, and $\gamma$ is set to $2$.

  

 \subsection{Optimization}
 We use Adam optimizer with initial learning rate of $ \alpha_{0} = 1e-4$ and progressively decrease it according to:
 \begin{equation}
 \label{eq:learningrate}
 \alpha = \alpha_{0} *\left(1-\frac{e}{N_{e}}\right)^{0.9}  
 \end{equation}
 where $e$ is an epoch counter, and $N_{e}$ is a total number of epochs (300 in our case). We draw input images in random order (ensuring that each training image is drawn once per epoch). 
 
 \subsection{Regularization}
 We use L2 norm regularization on the convolutional kernel parameters with a weight of $1e-5$.  We also use the spatial dropout with a rate of $0.2$  after the initial encoder convolution.
 
  \subsection{Data preprocessing and augmentation}
  We normalize all input images to have zero mean and unit std (based on non-zero voxels only). We  apply a random (per channel) intensity  shift ($-0.1..0.1$  of image std) and scale ($0.9..1.1$) on input image channels.  We also apply a random axis mirror flip (for all 3 axes) with a probability $0.5$. 

 \section{Results}
 \label{sec:results}
 
 We implemented our network in PyTorch~\footnote{https://pytorch.org/} and trained it on NVIDIA Tesla V100 32GB GPUs using BraTS 2019 training dataset (335 cases) without any additional in-house data. During training we used a random crop of size 160x192x128, which ensures that most image content remains within the crop area. We concatenated  4 available 3D MRI modalities into the 4 channel image as an input. The output of the network is 3 nested tumor subregions (after the sigmoid).
 
 We report the results of our approach  on BraTS 2019 validation (125 cases). We uploaded our segmentation results to the BraTS 2019 server for evaluation of per class dice, sensitivity, specificity and Hausdorff distances.

 The results of our model on the BratTS 2019 data are shown in Table~\ref{tab:valid} for the validation dataset and in Table~\ref{tab:testing} for the testing dataset.

\begin{table}
	\centering
	\caption{BraTS 2019 validation dataset results. Mean Dice and Hausdorff measurements of the proposed segmentation method. EN - enhancing tumor core, WT - whole tumor, TC - tumor core.}
	\label{tab:valid}
	\begin{tabular}{l|c|c|c|c|c|c}
		\hline
		& \multicolumn{3}{c|}{Dice} & \multicolumn{3}{c}{Hausdorff (mm)}  \\ \hline
		Validation dataset & ET & WT & TC & ET & WT & TC \\ \hline
		Single Model (batch 8) & 0.800 & 0.894 & 0.834 & 3.921 & 5.89 & 6.562 \\
		\hline
	\end{tabular}
\end{table}

\begin{table}

\centering
	\caption{BraTS 2019 testing dataset results. Mean Dice and Hausdorff measurements of the proposed segmentation method. EN - enhancing tumor core, WT - whole tumor, TC - tumor core.}
	\label{tab:testing}
	\begin{tabular}{l|c|c|c|c|c|c}
		\hline
		& \multicolumn{3}{c|}{Dice} & \multicolumn{3}{c}{Hausdorff (mm)}  \\ \hline
		Testing dataset & ET & WT & TC & ET & WT & TC \\ \hline
		Ensemble & 0.826 & 0.882 & 0.837 & 2.203 & 4.713 & 3.968 \\
		\hline
	\end{tabular}
\end{table}
 




Time-wise, each training epoch (335 cases) on a single GPU (NVIDIA Tesla V100 32GB) takes ~10min. Training the model for 300 epochs takes ~2 days. We trained the model on NVIDIA DGX-1 server (that includes 8 V100 GPUs  interconnected with NVLink); this allowed to train the model in ~8 hours. The inference time is 0.4 sec for a single model on a single V100 GPU.

\section{Discussion and Conclusion}
 \label{sec:conclusion}
In this work, we described a semantic segmentation network for brain tumor segmentation from multimodal 3D MRIs for BraTS 2019 challenge. We have experimented with various normalization functions, and found groupnorm and instancenorm to  perform equivalent, whereas batchnorm was always inferior, which could be due the fact of the largest batch size attempted being only 16. Since instancenorm is simpler to understand and implement, we used it for normalization by default. Multi-gpu systems, such as DGX-1 server, contains 8 GPU, which allows data-parallel implementation of batch size of 1 (where each each GPU get a batch of 1). We found the performance of multi-gpu system to be equivalent to a single gpu (batch 1) case, thus we used a batch of 8 by default, since it is almost 8 times faster to train.   We have also experimented with more sophisticated data augmentation techniques, including random histogram matching, affine image transforms, rotations, random image filtering, which did not demonstrate any additional improvements.   Increasing the network depth further did not improve the performance, but increasing the network width (the number of features/filters) consistently improved the results. Our BraTS 2019 final testing dataset results were 0.826, 0.882 and 0.837 average dice for enhanced tumor core, whole tumor and tumor core, respectively.

\bibliographystyle{splncs04}
\bibliography{am_brats2019}

\end{document}